# COGNITIVE TRANSFER OUTCOMES FOR A SIMULATION-BASED INTRODUCTORY STATISTICS CURRICULUM

MATTHEW D. BECKMAN
*Pennsylvania State University*
mdb268@psu.edu

ROBERT C. DELMAS
*University of Minnesota*
delma001@umn.edu

JOAN GARFIELD
*University of Minnesota*
jbg@umn.edu

**ABSTRACT**

*Cognitive transfer is the ability to apply learned skills and knowledge to new applications and contexts. This investigation evaluates cognitive transfer outcomes for a tertiary-level introductory statistics course using the CATALST curriculum, which exclusively used simulation-based methods to develop foundations of statistical inference. A common assessment instrument administered at the end of each course measured learning outcomes for students. CATALST students showed evidence of both near and far transfer outcomes while scoring as high, or higher on the assessed learning objectives, when compared with peers enrolled in similar courses that emphasized parametric inferential methods (e.g. the t-test).*

## 1. INTRODUCTION

Many students only complete one statistics course during their academic career, and consequently, their future use of statistics may depend heavily upon that limited exposure (Horton, 2015; Meng, 2011). As such, the statistics education community carries a responsibility to set forth the most productive and effective introductory curriculum possible. Ben-Zvi and Garfield, however, argued that "traditional approaches to teaching statistics focus on skills, procedures, and computations, which do not lead students to reason or think statistically" despite good performance in statistics courses (2005, p. 3). To this end, it is essential that the curriculum be optimized to promote the key ideals of statistical inference and probabilistic reasoning while preserving a flexibility that allows students to effectively apply these principles beyond the classroom in applications they encounter as students, professionals, citizens, and most any other domain in which information is aggregated and evaluated (Garfield, delMas, & Zieffler, 2012).

The ability to apply the skills and knowledge students learn in a course to novel problems is sometimes referred to as *cognitive transfer*. This transfer of learning outcomes is associated with the challenge of solving new problems based on prior experiences, especially when there has not been (or cannot be) explicit training in the context of the new task (Cox, 1997; Singley & Anderson, 1989). The premise of cognitive transfer is that functional understanding is more readily produced through broad understanding than by training a person to perform specific tasks (Bransford, Brown, & Cocking, 2000).

In order to study the ability of students to transfer what they have learned in an introductory statistics course, achievement of three groups of students enrolled in two different types of courses was compared on items that were designated as measuring different cognitive transfer outcomes. The two types of courses differed with respect to the emphasis or lack of emphasis on simulation-based methods to teach inference.

## 2. REVIEW OF THE LITERATURE

### 2.1. IMPORTANCE OF STATISTICAL THINKING

The ability to engage in statistical thinking, reasoning, and literacy are critical outcomes of the introductory statistics curriculum (delMas, 2002; Shaughnessy, 2007). Traditionally, a battery of procedures based on Normal and $t$-distributions have been the tools of choice to serve these goals, and Richard Scheaffer notes a general consensus on the content of the introductory statistics curriculum at the time (see discussion of Moore 1997). Wild, Pfannkuch, Regan, and Horton (2011), later contended that the traditional curriculum underestimates the complexity of its content and overestimates the capacity of students to surmount this complexity. Students are expected to process and utilize too many concepts for them to manage, resulting in an erosion of comprehension (Wild et al., 2011). Even with the multitude of topics to which students are exposed, other procedures with wide appeal and broad application such as ANOVA and basic nonparametric methods are often crowded out of the first course (Efron, 2000; Giesbrecht, Sell, Scialfa, Sandals, & Ehlers, 1997).

Others have raised similar concerns and have begun to develop and study strategies with hopes of reforming the introductory statistics curriculum. One theme that has emerged among researchers is a reduction of emphasis on traditional procedures rooted in Normal and $t$-distribution theory by supplementing or replacing them with simulation-based nonparametric methods (Ernst, 2004; Garfield et al., 2012; Tintle, VanderStoep, Holmes, Quisenberry, & Swanson, 2011; Wild et al., 2011).

### 2.2. COGNITIVE TRANSFER

Within research on the ability to transfer prior knowledge to new situations, a distinction is made is between *near* and *far* transfer (Bransford et al., 2000; Perkins & Salomon, 1988; Singley & Anderson, 1989). Near transfer can take place among contexts that are highly similar, whereas far transfer can occur across contexts that differ in superficial attributes yet build upon common concepts (Alexander & Murphy, 1999; Bransford et al., 2000; Cox, 1997; Paas, 1992). Although such attributes are generally discussed consistently in the literature, there is some room for interpretation regarding how *near* is near transfer or how *far* is far transfer. For example, Paas (1992) discusses an experiment conducted to evaluate factors influencing near and far transfer among classroom exercises, whereas Bransford et al. (2000) tended to refer to contexts outside of the classroom or school setting as involving far transfer. *Near* and *far* are relative terms in the transfer literature. Consequently, it is important to define each of the terms as they relate to this work. For the purpose of this study, the distinction between near and far transfer tasks is based on whether or not the course includes direct instruction related to the task (see Section 3.1).

Conditions needed to develop cognitive transfer have been documented in the research literature. Regardless of the distance or direction of transfer intended, intentional effort on the part of the student is necessary for successful development of the ability to transfer (Bransford et al., 2000; Perkins & Salomon, 1988; Singley & Anderson, 1989). In fact, students without explicit intervention will struggle or fail to transfer even when problem sets are extremely similar to course materials (Butterfield & Nelson, 1991; Cooper & Sweller, 1987; Lovett & Greenhouse, 2000; Reed, Dempster, & Ettinger, 1985; Singley & Anderson, 1989). In the context of learning introductory statistics, Garfield (2002) explained that statistics instructors often lay the groundwork of concepts and procedures and expect students to develop statistical reasoning or thinking through opportunities to apply content with software and data sets, but it seems this is simply not enough. Without further coaxing, most students do not abstract and generalize content effectively enough to achieve the cognitive plasticity required to assimilate novel or advanced applications (Garfield et al., 2012; Lovett & Greenhouse, 2000). Perkins and Salomon (1988) added that an important part of developing successful transfer outcomes depends upon the instructor training students to be intentional about learning with transfer in mind as opposed to merely introducing skills and concepts. The desired integrated understanding cannot be received passively from an instructor; it must be constructed by the learners for themselves (Broers, Mur, & Bude, 2004).

Abstraction of cognitive elements (i.e., knowledge, skills, and combinations thereof) seems to have a compelling role in preparing students for successful transfer outcomes. Essential to this end is the development and mastery of a rich schema for the content area, yet building and mastering schema is no small task (Bransford et al., 2000; Cooper & Sweller, 1987; Helfenstein, 2005; Lovett & Greenhouse, 2000; Paas, 1992; Perkins & Salomon, 1988). A schema tends to start small with a few similar problems, and then grows organically as the elements of the schema are repeatedly accessed and strengthened (Cooper & Sweller, 1987; Lovett & Greenhouse, 2000). In time, the boundaries of a schema domain swell and overlap with other domains such that the problem solver becomes increasingly equipped to assimilate a new problem into existing schema because of a depth and breadth of associated content mastery (Cooper & Sweller, 1987). In fact, some believe that higher ability students distinguish themselves by virtue of achieving greater abstraction and the facility to call on more distant connections (Goska & Ackerman, 1996).

Cognitive psychologists have frequently drawn upon a metaphor of computer processing speed and capacity in order to characterize elements of cognitive function (Deary, 2001). In essence, a person is constrained by some limited amount of cognitive capacity, and cognitive load describes the amount of burden imposed on the individual to process simultaneous demands (Cooper & Sweller, 1987; Lovett & Greenhouse, 2000). Although cognitive load is often discussed as a holistic concept, Sweller (1994) articulated a useful partition that distinguishes intrinsic and extraneous cognitive load.

Sweller (1994) described intrinsic cognitive load as the genuine burden of learning attributable to the nature of the content over which instructors have little or no control. If cognitive elements can be learned in succession, the intrinsic cognitive load is reduced because the need for interactivity among elements is removed (Sweller, 1994). It is often the case, however, that cognitive elements must be developed simultaneously because the most important outcomes lie in their interaction (Sweller, 1994). Some areas require greater element interactivity than others, and this fact is largely a fundamental truth of the content area and cannot be greatly influenced by instruction or environment (Singley & Anderson, 1989; Sweller, 1994). Statistics has been implicated as a high cognitive load domain (Paas, 1992). For example, students may be required to simultaneously grasp the abstract notions of an unseen population distribution, a random sample intended to represent the population, and a sampling distribution that characterizes the variability of a sample statistic, all with a goal of formulating a test or estimate of some quantity relevant to the original (and still unseen) population distribution.

Extraneous cognitive load can be described as an artificial burden directly attributed to the instructional methods (Sweller, 1994). If intrinsic cognitive load is relatively constant for a content area, the instructor has a responsibility to mitigate extraneous cognitive load by providing students the greatest opportunity for efficient learning outcomes. Suboptimal instruction imposes inefficient demand on resulting cognitive processing and hinders the potential for successful transfer (Paas, 1992; Singley & Anderson, 1989; Sweller, 1994). This paradigm frequently presents as a tendency to cover too much content too quickly for students to adequately process, resulting in impaired learning and transfer because students perceive the content as a set of disjointed facts that have not been organized and assimilated into usable schema (Bransford et al., 2000). Wild et al. (2011) and Garfield (1995) have echoed this phenomenon in statistics education, and suggested that many instructors are likely out of touch with the scale on which this problem affects their students.

## 2.3. SIMULATION-BASED INFERENCE METHODS FOR THE INTRODUCTORY STATISTICS CURRICULUM

As modern statistical methods become better understood and the practice of statistics changes, it seems prudent to periodically re-evaluate the teaching of this content, even at the introductory level. As courses change, it is important to examine if the new curricula are better at achieving important learning goals, such as transfer of learning. Many have argued for the use of nonparametric statistical methods (e.g. rank-based procedures) among the introductory statistics curriculum topics asserting that they are easier to learn and conceptually simpler than classical Normal theory procedures (Coakley, 1996; Hollander & Wolfe, 1999; Lehmann, 2009; Noether, 1980; Siegel, 1957). The same has been said of simulation-based nonparametric methods (e.g. bootstrap and randomization procedures), citing that they align well with intuition and build in a logical progression that is quite accessible to the introductory student (Cobb, 2007; Efron, 2000; Simon, Atkinson, & Shevokas, 1976; Tintle, Topliff, Vanderstoep, Holmes, & Swanson, 2012). Research by Gigerenzer & Hoffrage (1995) showed evidence that reasoning built upon intuition for frequencies (i.e., 43 times out of 1000) facilitates more successful learning outcomes than similar tasks phrased as probabilities (i.e., 0.043). As simulation-based methods connect naturally and tangibly to schema built upon frequency concepts (e.g., Ernst, 2004; Rossman, 2008), they represent a possible advantage over methods based on the Normal and *t*-Distributions that require more abstract probabilistic outcomes.

Moreover, the mathematics required are largely combinatorial in nature and often rely on little more than simple arithmetic, counting algorithms, and indicator functions (Coakley, 1996). For those procedures with more computational intensity like permutation tests, the difficulty of computation is generally related to combinatorial challenges such as collecting and cataloging all possible permutations of a data set rather than mathematical sophistication (Efron & Tibshirani, 1993; Ernst, 2004). In such cases, it is widely acceptable to use Monte Carlo (simulation and resampling) methods to generate a large number of resampled combinatorial outcomes, even if this deviation results in some outcomes counted more than once and others not at all (Efron & Tibshirani, 1993; Ernst, 2004).

Before modern computing, statisticians were forced to choose between mathematical approximations based on theoretical assumptions and a potentially untenable volume of trivial computations (Cobb, 2007; Mills, 2002). Yet, even Fisher, writing about permutation and randomization procedures, stated that "the statistician does not carry out this very simple and very tedious process, but his conclusions have no justification beyond the fact that they agree with those which could have been arrived at by this elementary method" (1936, p. 56). Because modern computing now permits previously unimaginable volumes of computations (Efron, 2000; Wild & Pfannkuch, 1999), the proper facility to incorporate simulation-based methods is now available to the introductory curriculum (Cobb, 2007; Mills, 2002).

The robustness of simulation-based inference techniques is due in part to the portability of relatively simple principles that generalize well to many situations (Cobb, 2007; Efron, 2000; Garfield et al., 2012). Garthwaite (1996) and Manly (1991) add that simulation-based procedures rely on less stringent distributional assumptions than parametric competitors, often with minimal loss of statistical power.

Simulation-based inference procedures can make concepts like sampling variability and sampling distributions quite lucid for the student (Cobb, 2007; Rossman, 2008; Tintle et al., 2011). Rather than imagine the comprehensive sampling distribution of some statistic if it were possible to obtain all possible samples from a potentially immeasurable population, students simulate a large number of possible samples and build an approximation of the sampling distribution directly. By constructing a permutation test distribution that represents an exact distribution of every possible outcome (or even approximating one through simulation), students should be able to easily deduce the proper interpretation of the $p$-value (Rossman, 2008). As curricula predicated on simulation-based methods would seem to promote such benefits, there is reason to compare simulation-based curricula and more traditional parametric curricula as equitably as possible in order to investigate potential strengths and limitations of each approach.

## 2.4. THE CATALST CURRICULUM

One approach to introductory statistics using simulation-based inference methods is the CATALST (Change Agents for Teaching and Learning Statistics) course (Garfield et al., 2012). This curriculum was designed to foster statistical thinking through methods such as bootstrapping and randomization testing. The CATALST curriculum built upon the call by Cobb (2007, p. 12) and others to utilize simulation-based methods to make explicit the *core logic of inference*. The curriculum as it was taught at the time of the study was organized into three modules as described by Garfield et al. (2012).

The first module of the CATALST course introduced students to random variables and simulation principles. The module began with a model eliciting activity (MEA—see, for example, Lesh & Doerr, 2003; Lesh, Hoover, Hole, Kelly, & Post, 2000) in which students studied randomness of playlists produced by an iPod Shuffle and developed algorithms to identify non-random playlists. Garfield et al. (2012) suggested that a major utility of beginning with an MEA is to present students with an opportunity to develop prior knowledge and build schema for statistical and probabilistic concepts within a familiar context. Students honed their understanding of chance models and simulation methods throughout the module, and then revisited the iPod Shuffle task to apply their learned skill set and segue to use of $p$-values to quantify evidence against a model (Garfield et al., 2012).

The second module introduced students to group comparisons. Students again began with an MEA motivated by a media article to compare reliability of airline arrival and departure times based on real data. Students created informal models for comparison of multiple airlines, then the class participated in another activity to develop schema for the characterization of distributions. Next, the class was introduced to randomization test procedures for quantitative and categorical variables. As the module progressed, the class became increasingly able to discuss concepts like variability and $p$-values in more formal terms. Also, design principles related to sampling, random assignment, sensitivity and specificity were considered. Students again revisited the opening MEA at the end of the module and attempted to produce solutions using methods learned during the module.

The third and final module led students through discussions about sampling distributions and confidence intervals using the nonparametric bootstrap. Students learned about fundamentals of parameter estimation including bias, precision, and the impact of sample size. The third module also formalized standard deviation as a measure of dispersion. The module concluded with discussion of Normal and *t*-Distribution methods and their parallels to the simulation-based inference methods discussed throughout the course.

## 2.5. THE NEED TO EXAMINE COGNITIVE TRANSFER IN INTRODUCTORY STATISTICS CLASSES

There is a general belief that an effective curriculum should promote successful cognitive transfer (Sternberg, 1998). The expectation is that students are likely to perform better with near transfer tasks because they more closely resemble the learned context, but the nature of disparity between near transfer performance and far transfer performance may indicate how well students have abstracted the cognitive elements of a curriculum.

A comparison of transfer outcomes between students exposed to the CATALST curriculum and those exposed to non-simulation-based introductory courses was conducted to evaluate the effectiveness of the simulation-based inference curriculum. The simulation-based curriculum studied here, and the assessment instrument used, were both developed as part of the CATALST project funded by a grant from the United States Government's National Science Foundation (NSF). It is also pertinent to note that no common or standardized curriculum was imposed for the non-simulation-based (i.e., non-CATALST) courses; instructors were expected to implement the curriculum that was in use for their course at the time of the study.

The objective of this study is to evaluate outcomes of cognitive transfer for two groups of students exposed to a CATALST curriculum, while also comparing their results to similar students exposed to non-simulation-based curricula. The cognitive transfer literature suggests that success of transfer outcomes largely depends on the ability of students to abstract relevant cognitive elements in order to apply their knowledge to a new scenario. The consistent use of simulation-based machinery in the CATALST curriculum gives students repeated exposure to key cognitive elements throughout the semester that may aid them in the required abstraction and generalization of these principles to novel tasks. Thus, it is expected that overall performance by CATALST students on near transfer and far transfer tasks would be at least as proficient as the comparison group and, furthermore, that performance on near transfer tasks would be at least as strong and perhaps even exceed performance on far transfer tasks because distance of transfer is defined based on similarity to exposures during instruction.

Therefore, the primary research question follows: How well do students exposed to a simulation-based curriculum accomplish near and far transfer upon completion of the course?

## 3. METHODS

In order to address the research question of this paper, cognitive transfer outcomes were compared between students enrolled in courses based on the simulation-based CATALST curriculum and students enrolled in non-simulation-based (i.e., Non-CATALST) courses using a common set of assessment items that were administered to students enrolled in both types of curricula.

### 3.1. THE ASSESSMENT INSTRUMENT

The assessment items were part of an instrument that was being developed in 2011 as part of the NSF-funded e-ATLAS project in concert with the CATALST project (Garfield et al., 2012). Two versions of the instrument were created, one for simulation-based courses such as the CATALST curriculum and one for non-simulation-based courses. Although each version of the instrument consisted of 27 forced-choice items, only a subset of 21 items that was common to both instruments was used for this study. Of these 21 items, 16 (76%) were modified versions of items from the Comprehensive Assessment of Outcomes in Statistics (CAOS) test (delMas, Garfield, Ooms, & Chance, 2007) and assessed the same learning outcomes designated for the corresponding items on the CAOS test. The five newly developed items (see Table 1) assessed learning outcomes related to representative samples (item 3), understanding null hypotheses (items 15 and 16), and understanding statistical significance (items 20 and 21).

Table 1 lists the learning outcome assessed by each of the 21 common items. Several characteristics of the 21 common items are represented in Table 1. First, some of the items shared a common context, forming what is sometimes referred to as a testlet (Wainer, Sireci & Thissen, 1991). Each set of items that shares a common context is enclosed within a box in Table 1 (e.g., items 8 and 9; items 11 through 14). Individual items that were not part of a testlet were scored 0 = incorrect, 1 = correct, whereas a single score was computed for a testlet instead of for each individual item within a testlet. For most of the testlets, the single score equaled the proportion of items in the testlet that were answered correctly (e.g., possible scores for the testlet consisting of items 11 through 14 were 0, 0.25, 0.50, 0.75 and 1). The testlet consisting of items 8 and 9, however, was scored using a different method. To answer this testlet correctly, the respondent had to make a correct choice for item 8 and select a correct reason for the item 8 choice in item 9 (1 = both items correct, 0 = otherwise).

Second, Table 1 classifies each of the test items as either near or far transfer based on the content of the CATALST curriculum. Tasks were classified according to the type of exposure that students received throughout the curriculum for the predominant concept evaluated by that item as shown in Table 1, which was adapted from Sabbag (2012). The three types of exposure considered for estimating transfer distance include direct instruction, indirect instruction, and practice. Near transfer tasks require that students have been exposed to some amount of direct instruction on the concept in addition to any amount of indirect instruction or practice. Alternatively, far transfer tasks correspond to concepts for which students could have received indirect instruction and/or practice, but no direct instruction.

Direct instruction indicates that students were provided experiences or definitions that lead them to an understanding of the concept, which was then specifically identified. By contrast, indirect instruction indicates that the concept was not mentioned during instruction, but students were provided experiences designed to build their understanding. Finally, concepts presented during practice assumed that students already had some understanding of the concept, which they provided without additional direct or indirect instruction.

*Table 1: Measured learning goal and frequency of exposure type (i.e., number of lessons) within the CATALST curriculum for each assessment item\**

|      |                                                                                                                     | CATALST Exposure Type |          |          |          |
|------|---------------------------------------------------------------------------------------------------------------------|-----------------------|----------|----------|----------|
| Item | Measured Learning Goal                                                                                              | Direct                | Indirect | Practice | Transfer |
| 1    | Purpose of random assignment                                                                                        | 3                     | 3        | 3        | Near     |
| 2    | Factors that allow a sample of data to be generalized to the population                                             | 3                     | -        | 4        | Near     |
| 3    | Factors that allow a sample of data to be representative of the population                                          | 3                     | -        | 4        | Near     |
| 4    | Correlation does not imply causation                                                                                | -                     | 5        | 2        | Far      |
| 5    | Match a scatterplot to a verbal description of a bivariate relationship                                             | -                     | -        | -        | Far      |
| 6    | Statistics from small samples vary more than statistics from large samples                                          | 1                     | 5        | -        | Near     |
| 7    | Expected patterns in sampling variability                                                                           | 2                     | 3        | -        | Near     |
| 8    | Meaning of variability in the context of repeated measurements and in a context where small variability is desired  | -                     | 5        | -        | Far      |
| 9    | Meaning of variability in the context of repeated measurements and in a context where small variability is desired  | -                     | 5        | -        | Far      |
| 10   | Using characteristics of distributions to compare two groups                                                        | 2                     | -        | -        | Near     |
| 11   | Detect a misinterpretation of a confidence level (the range of observed values)                                     | -                     | 4        | -        | Far      |
| 12   | Correct interpretation of confidence interval                                                                       | 1                     | 1        | 3        | Near     |
| 13   | Detect a misinterpretation of a confidence level (percentage of population data values between confidence limits)   | -                     | 4        | -        | Far      |
| 14   | Correct interpretation of confidence interval                                                                       | 1                     | 1        | 3        | Near     |
| 15   | Understanding of what the null hypothesis represents                                                                | 6                     | 2        | 5        | Near     |
| 16   | Relationship between the $p$-value and the null hypothesis                                                          | 4                     | 4        | 5        | Near     |
| 17   | Recognize an incorrect interpretation of a $p$-value (probability of a treatment being less effective)              | 1                     | 3        | -        | Near     |
| 18   | Recognize an incorrect interpretation of a $p$-value (probability of a treatment being more effective)              | -                     | 4        | -        | Far      |
| 19   | Correct interpretation of a $p$-value                                                                               | -                     | 4        | -        | Far      |
| 20   | Recognize a misinterpretation of a statistically significant result (referring to incorrect parameter)              | -                     | 8        | -        | Far      |
| 21   | Recognize a misinterpretation of a statistically significant result (not discerning from practical significance)    | -                     | 9        | -        | Far      |

*Items framed within a box form a testlet

Table 1 shows that 11 items (1, 2, 3, 6, 7, 10, 12, 14, 15, 16, 17) met the definition of near transfer tasks, while 10 items (4, 5, 8, 9, 11, 13, 18, 19, 20, 21) met the definition of far transfer tasks. When a testlet consisted of both near and far items, two testlet scores were computed, one indicating the proportion of correct responses for the near transfer items in a testlet and the other indicating the proportion of correct responses to the far transfer items. This resulted in seven individual item and two teslet scores for measuring near transfer, and two individual item and four testlet scores for measuring far transfer. An overall near transfer score and an analogous far transfer score were computed as the average proportion correct of the corresponding set of items and testlet scores for each transfer type.

**3.2. DATA COLLECTION**

The data used for this study are secondary data originally collected for evaluation of the CATALST Project (Garfield et al., 2012). After reviewing the CATALST Project and learning of the assessment data that had been collected, the first author contacted the CATALST principal investigator and asked permission to use the data to conduct a secondary analysis of transfer outcomes.

Students from eight different institutions participated in the study. The non-simulation-based version of the assessment instrument was administered to 440 students from 10 class sections among seven institutions around the United States during fall 2011 and spring 2012. The simulation-based version of the assessment instrument was administered to 289 students from nine class sections among six institutions (four of the classes were at the University of Minnesota) during spring 2012. Five of the institutions had both CATALST and non-CATALST sections of introductory statistics that administered the respective versions of the assessment instrument.

Those instructors selected for participation in the CATALST Project were provided with lesson plans and materials as well as regular access to the CATALST project team via telephone discussions every two weeks and other communications. Instructors were granted the liberty to administer the instrument as they saw fit, and some instructors (4 of 9) reported using the assessment instrument as a final exam for the CATALST course. Participating instructors were also instrumental in administering the assessment instrument to students enrolled in a course that did not teach simulation-based inference methods. In some cases, the CATALST instructors administered the assessment instrument to their own students enrolled in a non-CATALST course during a term prior to the instructor teaching the CATALST curriculum. Others recruited a colleague to administer the assessment instrument to students enrolled in a non-CATALST course. This process was intended to lend some assurance that students within the CATALST and non-CATALST sections would be comparable within an institution, although it is noteworthy that no common curriculum was imposed for the non-CATALST courses.

**3.3. ANALYSIS OF TRANSFER OUTCOMES AMONG SIMULATION-BASED CURRICULUM PARTICIPANTS**

For the purpose of this analysis, students who were taught using the CATALST curriculum were partitioned into two groups. The first group included students taught the CATALST curriculum from an instructor at the University of Minnesota (CAT-UMINN), and the other group included students taught the CATALST curriculum by an instructor not otherwise associated with the University of Minnesota (CAT-REPL). As faculty at the University of Minnesota largely developed the CATALST curriculum, the proximity of CAT-UMINN instructor group to the curriculum developers could influence their delivery of the curriculum in some way that was not accessible to those outside the University of Minnesota. Lastly, although the main focus of the study pertains to distance of transfer among students exposed to the CATALST curriculum, results for analogous items among non-CATALST curriculum (NON-CAT) participants were included as an approximate comparison group although actual distance of transfer is unknown among this group.

In order to address whether differences exist between learning outcomes for students exposed to a simulation-based curriculum and students exposed to a non-simulation-based curriculum as measured by the assessment instrument, the three groups were compared on overall scores for the common set of assessment items. Moreover, the primary research question, "How well do students exposed to a simulation-based curriculum accomplish near and far transfer upon completion of the course?," can be addressed by assessing whether or not there is an interaction between curriculum group and type of transfer. Students were not randomly assigned to the different curriculum groups and the different courses were delivered to entire sections of students. Therefore, statistical analyses were based on a linear mixed effects (LME) model where students were clustered within class section. Two contrasts were defined for the curriculum group, one for the difference between the CAT-UMINN and CAT-REPL groups and the other for the mean difference between the CAT-REPL and NON-CAT groups. These two contrasts allowed for an assessment of the effect of the proximity to the curriculum developers and the effect of curriculum within institution, respectively. Analysis of the LME model was performed in R using the lme4 package (Bates, Mächler, Bolker & Walker, 2015a, b) to estimate model parameters and the lmerTest package (Kuznetsova, Brockhoff & Christensen, 2016) to produce estimates of standard errors, p-values, and confidence intervals. The lsmeans() function from the lmerTest package was used to produce least squares means and confidence interval estimates. The lsmeans() function uses a Satterthwaite approximation to the degrees of freedom, which results in a conservative estimate of *p*-values and confidence intervals (Kuznetsova, Brockhoff & Christensen, 2016).

## 4. RESULTS

### 4.1. TRANSFER OUTCOMES AMONG CURRICULUM GROUPS

An initial LME model was fitted with proportion correct as the dependent variable, fixed effects for both curriculum group and item transfer type (near or far), the interaction between the two fixed effects, and a random effect for students clustered within class section. Traditionally, transfer item type can be considered to represent a within-subjects factor. A random effect for item transfer type varying within student, however, was not fit to the model because there are only two levels for the factor (near and far transfer) and the resulting model would be fully saturated (i.e., the number of observations and the number of random effects are equal). Consistent with the model being fully saturated, an LME model that included a random effect for item transfer type produced a near perfect correlation ($r = 0.98$) between the model fitted estimates and actual scores for the students. Therefore, a random effect for item transfer type varying within student was not added to the model and item transfer type was treated as a fixed effect, with near and far transfer items as the only two levels of interest. Several contrasts were specified to conduct specific comparisons for interpreting main and interaction effects. All contrasts were defined as the mean difference between two groups, with the first group assigned a contrast coefficient of 1 and the second group a contrast coefficient of -1. Two contrasts were specified for the fixed effect of the curriculum group factor. The first contrast compared the difference between the means of the CAT-UMINN and CAT-REPL groups and the second contrast compared the CAT-REPL group and the Non-CATALST curriculum group. A contrast was specified for the fixed effect of transfer item type as the difference between the mean proportion correct of the near and far items. Assessment of model assumptions did not indicate any noteworthy violations (see Appendix A). Table 2 shows a summary and parameter estimates for the resulting LME model.

*Table 2. Linear mixed effects model parameter estimates with confidence intervals*

| Random Effect | Estimate [95% CI] | | | |
|---|---|---|---|---|
| Class section standard deviation | 0.069 [0.044, 0.093] | | | |
| Student level standard deviation | 0.172 [0.166, 0.178] | | | |
| Fixed Effect | | df | $t$ | $p$ |
| Grand Mean [intercept] | 0.608 [0.574, 0.642] | | | |
| Curriculum group: UMINN - REPL | 0.083 [0.031, 0.135] | 15 | 3.030 | 0.008 |
| Curriculum group: REPL - NON-CAT | 0.079 [0.037, 0.121] | 15 | 3.541 | 0.003 |
| Transfer item type: Near - Far | 0.023 [0.013, 0.033] | 1435 | 4.485 | < 0.001 |
| UMINN - REPL by Transfer | 0.011 [-0.005, 0.026] | 1435 | 1.373 | 0.170 |
| REPL - NON-CAT by Transfer | 0.021 [0.009, 0.033] | 1435 | 3.380 | < 0.001 |

Results from the analysis of the fixed effects from the LME model show statistically significant effects for curriculum group, transfer type of assessment items, and the interaction between curriculum group and transfer type (see Table 3). Table 4 presents the estimated mean proportion correct on near and far transfer items, as well as the total set of assessment items, for students in each of the three curriculum groups. All of the means in Table 4 can be computed from the parameter estimates in Table 2 and the corresponding sets of coefficients for each contrast (see Appendix B). CAT-UMINN group had the highest estimated mean proportion correct, followed by the CAT-REPL students and then the NON-CAT group, for all three assessment measures. In addition, the estimated mean proportion correct was higher for near transfer items than for far transfer items for all three curriculum groups.

*Table 3. ANOVA results for fixed effects from the LME model*

| Fixed Effect | Degrees of Freedom | $F$ | $p$ |
|---|---|---|---|
| Curriculum Group | 2, 15.1 | 7.461 | 0.0056 |
| Transfer Type | 1, 1435.1 | 20.112 | < 0.0001 |
| Interaction | 2, 1435.1 | 5.713 | 0.0034 |

*Table 4. Mean proportion correct estimated from the LME model*

| | | Mean [95% CI] | | |
|---|---|---|---|---|
| | $n$ | Near Transfer | Far Transfer | All Items |
| CAT-UMINN | 138 | 0.725 [0.646, 0.804] | 0.657 [0.578, 0.736] | 0.691 [0.614, 0.768] |
| CAT-REPL | 151 | 0.636 [0.565, 0.708] | 0.570 [0.499, 0.642] | 0.604 [0.534, 0.673] |
| NON-CAT | 440 | 0.531 [0.482, 0.581] | 0.527 [0.477, 0.577] | 0.529 [0.480, 0.578] |
| All Students | 729 | 0.631 [0.592, 0.670] | 0.585 [0.546, 0.624] | 0.608 [0.570, 0.646] |

Contrast analyses were conducted to further understand the main effect of curriculum group and the interaction (see Table 2). The contrast comparing the CAT-UMINN and CAT-REPL groups was statistically significant, as was the contrast between the CAT-REPL group and the Non-CATALST curriculum group. While both of these contrasts are statistically significant, these results are not consistent with the estimated 95% confidence interval for each mean difference. Both confidence intervals based on all items for the mean difference between the respective groups include 0, although the lower limit of each confidence interval is near 0 in both cases (see Table 5).

*Table 5. Mean difference in proportion correct estimated from the LME model*

| Difference | Group or Measure | Mean Difference [95% CI] |
|---|---|---|
| Near - Far | | |
| | CAT-UMINN | 0.068  [0.027, 0.108] |
| | CAT-REPL | 0.066  [0.027, 0.105] |
| | NON-CAT | 0.005 [-0.018, 0.027] |
| | All Groups | 0.046  [0.026, 0.066] |
| CAT-UMINN - CAT-REPL | | |
| | Near | 0.089 [-0.018, 0.195] |
| | Far | 0.087 [-0.020, 0.193] |
| | All Items | 0.088 [-0.016, 0.191] |
| CAT-REPL - NON-CAT | | |
| | Near | 0.105  [0.018, 0.192] |
| | Far | 0.044 [-0.043, 0.131] |
| | All Items | 0.074 [-0.010, 0.159] |

With respect to the interaction between curriculum group and type of transfer item, there was not strong evidence of the interaction effect when the University of Minnesota CATALST curriculum group (CAT-UMINN) was compared to the students who experienced the CATALST curriculum at other institutions (CAT-REPL). This is consistent with the confidence intervals reported in Table 5 that indicated the difference between the CAT-UMINN and CAT-REPL groups is not statistically significant for both near and transfer items. There was strong evidence of the interaction effect when the CAT-REPL group was compared to the Non-CATALST curriculum group, since the CAT-REPL and Non-CATALST results appear to meaningfully differ among near transfer tasks but not among far transfer tasks. Inspection of the mean difference in proportion correct reported in Table 5 and the interaction plot in Figure 1 shows that the difference in mean proportion correct between near and far transfer items appeared similar for the CAT-UMINN and CAT-REPL groups, whereas there was virtually no difference between mean proportion correct for the two types of transfer items for the Non-CATALST curriculum group. Estimates of confidence intervals for the difference between mean proportion correct on the near and far transfer items for each curriculum group (Table 5) and a plot of mean proportion correct in Figure 1 corroborate the description of the interaction.

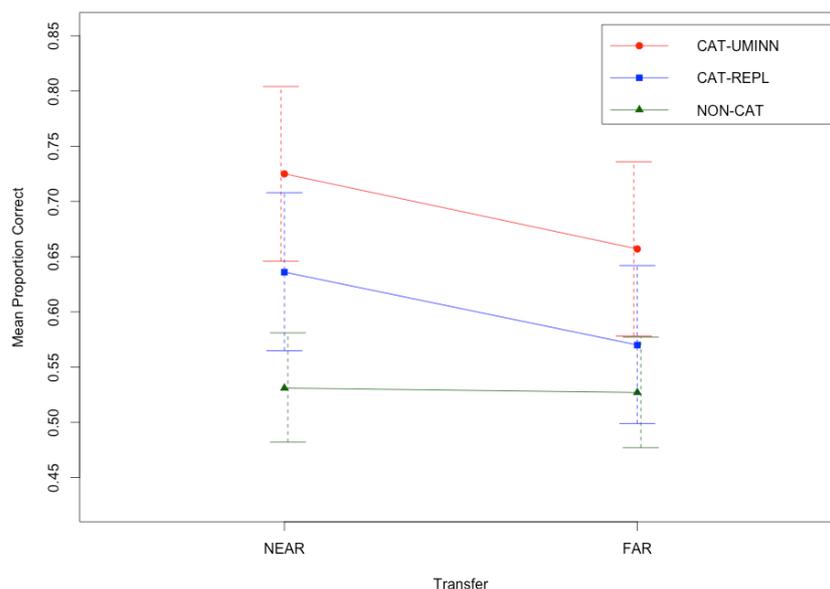

*Figure 1. Plot of estimated mean proportion correct for near and far items by curriculum group; vertical dashed bars represent estimated 95% confidence intervals.*

## 5. DISCUSSION

### 5.1. TRANSFER OUTCOMES AMONG SIMULATION AND NON-SIMULATION CURRICULUM GROUPS

This study was designed to evaluate the extent of transfer expected of students who took a CATALST course, either at the original development location (University of Minnesota) or at other schools. CATALST student results were compared to a third group of students, who enrolled in courses that did not extensively use simulation and modeling approaches. The data gathered provide evidence of successful near and far transfer outcomes among CATALST students as well as results that were superior overall to those produced by students of Non-CATALST courses. The performance of the Non-CATALST group on both the near and far transfer items is comparable to the overall performance of students on the CAOS test (delMas, 2014; delMas et al., 2007), whereas performance by the CATALST students is markedly higher. Recall that tasks were identified to be *far transfer* on the basis that they measure learning goals for which students in a CATALST curriculum did not receive any direct instruction (evaluated by Sabbag, 2012, and reproduced in Table 1). Because there was no similar parsing of instructional contact available for the Non-CATALST courses, students in Non-CATALST courses may have had the benefit of direct instruction for some or all of the learning outcomes assessed by those tasks given the general alignment of the assessment items with the CAOS test (delMas et al., 2007). If all else were assumed equivalent, this would likely create an advantage for Non-CATALST students, which would amplify the significance of a result suggesting that they were outperformed by CATALST students on those tasks.

It is also worth reiterating the extent to which students in the CATALST and Non-CATALST groups would be comparable to one another. Based on the process for recruiting instructors to participate in the study, in many cases the students in Non-CATALST courses were enrolled at the same institution and taught by the same instructor as a group of students included in CATALST group. Although the contrasts of interest would have greater statistical power if the analysis had adjusted for these commonalities, such adjustment was not feasible since several of the Non-CATALST courses were taught by a colleague at the same institution as a CATALST instructor, and one Non-CATALST course was taught without a CATALST counterpart.

One stated expectation had been that CATALST students at the University of Minnesota may outperform CATALST students at other institutions due to proximity of the curriculum developers concentrated at the University of Minnesota. This expectation was supported by the results through the statistically significant difference favoring the CAT-UMINN group compared to the CAT-REPL group, and the lack of a statistically significant interaction with item transfer type. We speculate that prior experience with both teaching and development of the CATALST curriculum by the University of Minnesota instructors is a key reason for the difference.

The statistical evidence indicates that CATALST students at institutions other than the University of Minnesota also outperformed the Non-CATALST students on near transfer items. The majority of the students in the two groups were at the same institutions. To the extent that the assessment instrument reflects desirable outcomes of the post-secondary introductory statistics curriculum, this would provide some measure of evidence that a simulation-based curriculum may accomplish these outcomes more effectively. In addition, the noticeably higher performance of the University of Minnesota CATALST students suggests that students at the other institutions may show higher performance as the instructors gain experience in teaching the CATALST curriculum, although additional training support to ensure curriculum fidelity might also be needed.

**5.2. LIMITATIONS**

One important limitation of this study is the acknowledgment that the so-called Non-CATALST courses did not represent a single curriculum. The distance of transfer for each item is dependent on the content of the specific curriculum to which a student is exposed. Also, each participating instructor was afforded the liberty to implement the assessment instrument for whatever purpose they deemed most appropriate. Such purposes may include as a final exam, study guide, for extra credit, homework, etc. This lack of rigorous controls limits the generalizability of comparisons made.

For the same reason that Non-CATALST students may have an advantage on items that measure far transfer outcomes for the CATALST students, the Non-CATALST students may have a disadvantage on items that measure near transfer outcomes for the CATALST students. That is, CATALST students are known to have benefited from direct instruction on all tasks with the near transfer designation in this paper, yet it could not be confirmed whether Non-CATALST students had direct instruction for all of the same learning outcomes. Some caution is therefore warranted when comparing performance of near transfer tasks. Arguably, the effect is tempered because the assessment instrument was based largely on the CAOS test, which was developed and established before the CATALST curriculum had been developed, so the content was designed to largely represent topics typical of an introductory (Non-CATALST) curriculum. Consequently, one might expect most, if not all, topics assessed as near transfer to be treated with direct instruction in any introductory statistics curriculum (CATALST and Non-CATALST alike). Upon inspection of learning goals associated with near transfer items as defined by Table 1, we expect many that a conventional post-secondary introductory statistics course would commonly include direct instruction on these topics.

The results of the CAT-UMINN group may be biased as the CATALST curriculum and assessment instrument were developed by the same research team. The instructors in the CAT-REPL group were not among the research team involved in development of the CATALST curriculum or the assessment instrument, so the results of that group should be insulated from such bias. Furthermore, since the CAT-REPL instructors were teaching the CATALST curriculum for the first time while participating in the project, they may have been less comfortable with the simulation-based approach. By contrast, Non-CATALST course instructors were invited to use any curriculum they wished, so it is reasonable to speculate that it would be one with which the instructor was quite comfortable. This could serve to disadvantage the CATALST students, but is noted here among the limitations of the study because the effect cannot be quantified and therefore obscures the comparisons of interest.

Lastly, instructors were not randomly selected for participation in the study, nor were students randomly assigned to each curriculum. Even if researchers originally involved in the CATALST Project attempted to represent a meaningful diversity of instructors and institutions, without random selection there is risk that an important subset of the desired population has not been adequately represented. Furthermore, without randomly assigning students to each curriculum we cannot confidently rule out that any comparison of outcomes between curricula would be meaningfully influenced by confounding variables.

**5.3. IMPLICATIONS FOR RESEARCH**

The results of this analysis complement findings that have suggested benefits of a simulation-based introductory curriculum for improved learning outcomes and retention (Maurer & Lock, 2016; Tintle et al., 2012), yet there is much work to be done to better understand transfer outcomes associated with a simulation-based curriculum like CATALST. Since cognitive transfer is a progression that develops over time, it is no small task to identify early signs of successful or unsuccessful transfer within the context of 15-week semesters. The divergence between simple performance on course content and the robust learning that facilitates successful transfer is likely quite subtle in this short amount of time, so further research is needed in order to study transfer outcomes of greater distance. This may include studying carryover effects to a second course in statistics among those students exposed to a simulation-based approach in a first course, or it may include studying statistical thinking among those students outside of the academic environment.

Lovett and Greenhouse (2000) describe an important link between transfer outcomes and cognitive load that deserves exploration in this context. Additional research is needed in order to begin studying the impact of the simulation-based curriculum on intrinsic and extraneous cognitive load. There is reason to speculate that a simulation-based curriculum can be tuned to reduce intrinsic cognitive load, as it relies on a consistent set of machinery that is flexible in application and interpretation across a number of contexts. Optimizing cognitive load is an important consideration for any subject matter if instructional practices are to mature and evolve.

**5.4. IMPLICATIONS FOR INSTRUCTION**

The results of this analysis and those of similar implementations of simulation-based curricula (for example, Tintle et al., 2011) have corroborated that students performed at least as well as those exposed to a more conventional introductory statistics curriculum. This evidence should continue to bolster the confidence of introductory statistics instructors to incorporate simulation-based methods within their own curricula. If complete adoption of a simulation-based curriculum is not feasible, applets or other tools that use a simulation-based approach to simply introduce inferential thinking could be of great benefit.

18Maurer, K., & Lock, D. (2016). Comparison of learning outcomes for simulation-based and traditional inference curricula in a designed educational experiment. *Technology Innovations in Statistics Education 9*(1).
    [Online: http://escholarship.org/uc/item/0wm523b0]
Meng, X. L. (2011). Statistics: Your chance for happiness (or misery). *The Harvard Undergraduate Research Journal, 2*(1), 21–27.
Mills, J. D. (2002). Using computer simulation methods to teach statistics: A review of the literature. *Journal of Statistics Education, 10*(1).
    [Online: http://www.amstat.org/publications/jse/v10n1/mills.html]
Moore, D. S. (1997). New Pedagogy and New Content: The Case of Statistics (with discussion), *International Statistical Review, 65*, 123-165.
Noether, G. E. (1980). The role of nonparametrics in the introductory statistics courses. *The American Statistician, 34*(1), 22–23.
Paas, F. G. W. C. (1992). Training strategies for attaining transfer of problem solving skill in statistics: A cognitive-load approach. *Journal of Educational Psychology, 84*(4), 429–434.
Perkins, D. N., & Salomon, G. (1988). Teaching for transfer. *Educational Leadership, 46*, 22–32.
Reed, S. K., Dempster, A., & Ettinger, M. (1985). Usefulness of analogous solutions for solving algebra word problems. *Journal of Experimental Psychology: Learning, Memory, and Cognition, 11*, 106–125.
Rossman, A. J. (2008). Reasoning about informal statistical inference: One statistician's view. *Statistics Education Research Journal, 7*(2), 5–19.
    [Online: http://iase-web.org/documents/SERJ/SERJ7(2)_Rossman.pdf]
Sabbag, A. (2012). *GOALS-CATALST Blueprint; frequency and type of content exposure in CATALST curriculum for GOALS-RAND items*. [Unpublished raw data.]
Shaughnessy, J. M. (2007). Research on statistics learning and reasoning. In F. K. Lester (Ed.), *Second handbook of research on mathematics teaching and learning* (pp. 957–1009). Charlotte, NC: Information Age Publishing and NCTM.
Siegel, S. (1957). Nonparametric statistics. *The American Statistician, 11*(3), 13–19.
Simon, J. L., Atkinson, D. T., & Shevokas, C. (1976). Probability and statistics: Experimental results of a radically different teaching method. *American Mathematical Monthly, 83*(9), 733–739.
Singley, M. K., & Anderson, J. R. (1989). *The transfer of cognitive skill*. Cambridge, MA: Harvard University Press.
Sternberg, R. J. (1998). Metacognition, abilities, and developing expertise: What makes an expert student? *Instructional Science, 26*(1), 127–140.
Sweller, J. (1994). Cognitive load theory, learning difficulty, and instructional design. *Learning and instruction, 4*(4), 295–312.
Tintle, N., Topliff, K., Vanderstoep, J., Holmes, V., & Swanson, T. (2012). Retention of statistical concepts in a preliminary randomization-based introductory statistics curriculum. *Statistics Education Research Journal, 11*(1), 21–40.
    [Online: http://iase-web.org/documents/SERJ/SERJ11(1)_Tintle.pdf]
Tintle, N., Vanderstoep, J., Holmes, V., Quisenberry, B., & Swanson, T. (2011). Development and assessment of a preliminary randomization-based introductory statistics curriculum. *Journal of Statistics Education, 19*(1).
    [Online: http://www.amstat.org/publications/jse/v19n1/tintle.pdf]
Wainer, H., Sireci, S. G., & Thissen, D. (1991). Differential testlet functioning: Definitions and detection. *Journal of Educational Measurement, 28*(3), 197–219.

MATTHEW D. BECKMAN
Pennsylvania State University
326 Thomas Building
University Park, PA 16802 USA

**APPENDIX A: ASSESSMENT OF MODEL ASSUMPTIONS**

Assessment of model assumptions did not indicate any noteworthy violations. Although the proportion correct measures for both near and far transfer items were discrete, the distributions of the dependent variable for each curriculum group did not appear to be extremely different from a true normal distribution (see Figure A1). A modified Levene's test for the assumption of homogeneity of variance was not statistically significant for the proportion correct on the near and far transfer items for the three groups of students ($F(5, 1452) = 1.624$, $p = 0.150$), and Box's M test of the equality of the covariance matrices was not statistically significant ($M(6) = 10.633$, $p = 0.100$).

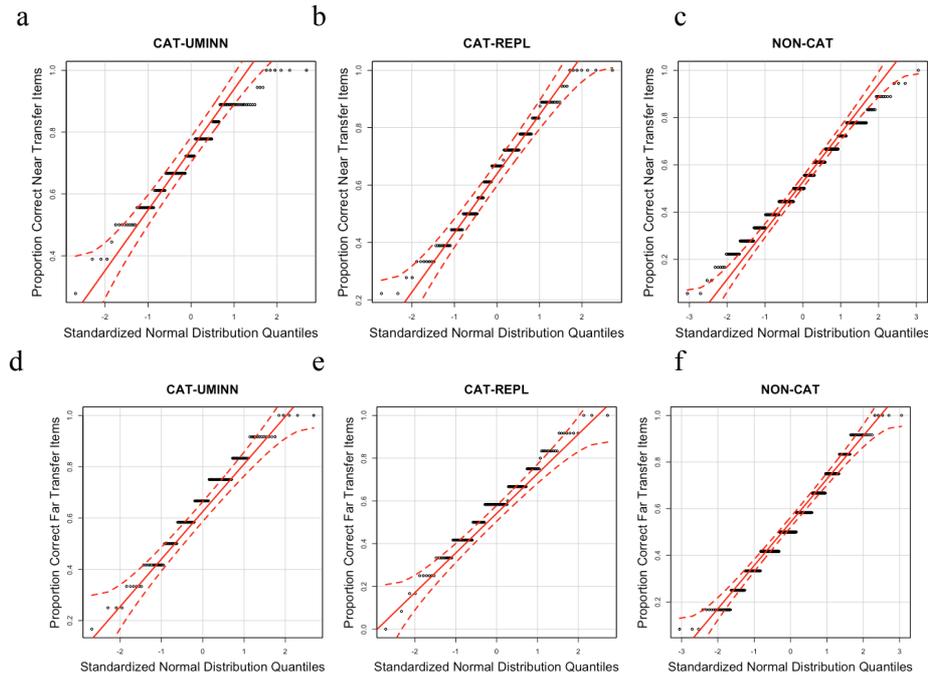

*Figure A1. Normal quantile-quantile plots with 95% confidence bands for near and far transfer scores by each curriculum group*

A null model that included only the random effect for students clustered within class section produced an intra-class correlation of 0.225, indicating that 22.5% of the variance in mean proportion correct scores was associated with variability among the 19 different class sections. A LOESS smoother with a 95% confidence band fitted to the residual plot indicated that a linear model is appropriate and the scatter of points did not indicate an extreme violation of the conditional homogeneity of variance assumption (see Figure A2). Confidence intervals for the random effects were relatively small (Table 2), indicating reliable estimation of both the between class and student level variances in the model.

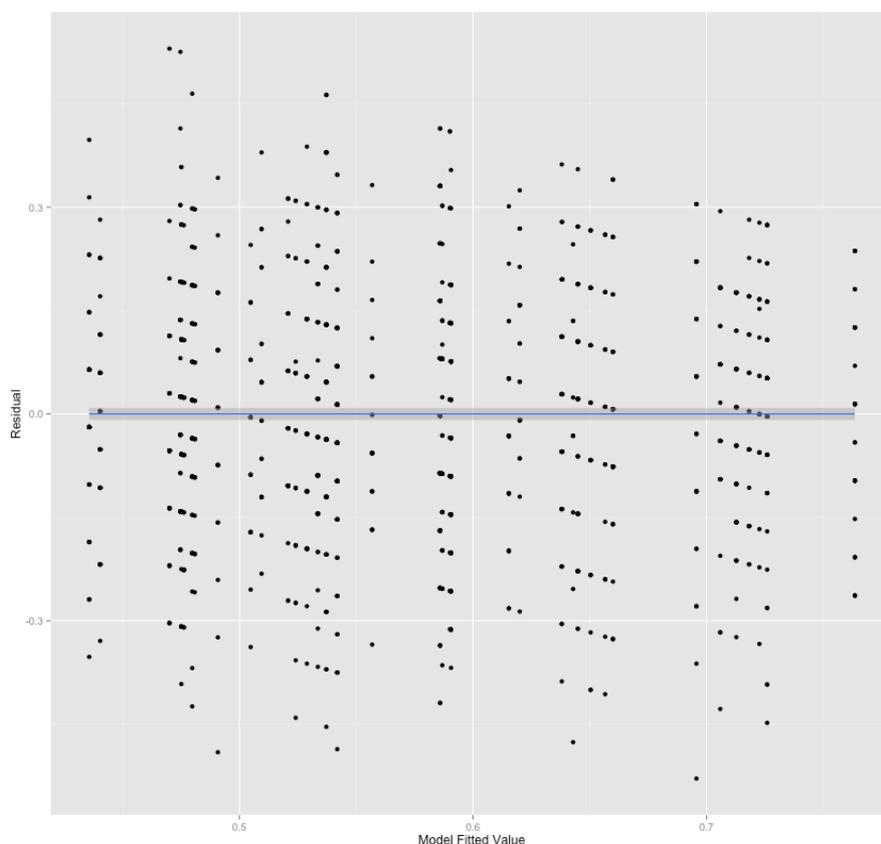

*Figure A2. Residual plots for the linear mixed effects models*

**APPENDIX B: RELATIONSHIP BETWEEN CONTRAST COEFFICIENTS AND MODEL PARAMETER ESTIMATES**

The model parameter estimates in Table 2 were estimated with respect to the contrast coefficients defined for each of the three contrasts. Each mean in Table 4 can be computed from the set of parameters and the corresponding contrast coefficients. The *y*-intercept is included in all computations of mean estimates. For the contrasts involving curriculum group, the coefficient is 0 when a group is not included in the contrast, or either 1 or -1 when the group is included (e.g., the coefficients for the Curriculum group: UMIN – REPL contrast are 1: CAT-UMINN, -1: CAT-REPL, 0: NON-CAT). For the contrast of the two transfer item types, the coefficient is 0 when all items are used to compute the mean. Otherwise, the coefficient is 1 when only near transfer items are included and -1 when only far transfer items are included. For each interaction contrast, the corresponding contrast coefficient for the main effect of curriculum group and the main effect of transfer item type are multiplied together to determine the weight. Each parameter estimate is then multiplied by its corresponding coefficient or product of coefficients, and the resulting products are summed to produce the mean estimate.

For example, the overall mean for all items and all students in Table 4 (0.608) is computed across all items and does not involve any of the contrasts. Consequently, the estimated mean is simply equal to the y-intercept reported in Table 2.

To compute the group mean on all items for only the CAT-UMINN group (0.691) involves only the *y*-intercept plus the coefficient for Curriculum group: UMINN – REPL.

The contrast coefficient for the CAT-UMINN group is 1 for this contrast. All other effects will have a contrast coefficient of 0. Therefore, the mean is computed as:

$(1)(0.608) + (1)(0.083) + (0)(0.079) + (0)(0.023) + (0)(0)(0.011) + (0)(0)(0.021) = 0.691$

Similarly, the group mean on all items for the CAT-REPL group is computed as:

$(1)(0.608) + (-1)(0.083) + (1)(0.079) + (0)(0.023) + (0)(0)(0.011) + (0)(0)(0.021) = 0.604$

because the CAT-REPL group has a coefficient of -1 for the contrast with the CAT-UMINN group and a coefficient of 1 for the contrast with the NON-CAT group.

The mean for near transfer items across all curriculum groups is computed as:

$(1)(0.608) + (0)(0.083) + (0)(0.079) + (1)(0.023) + (0)(0)(0.011) + (0)(0)(0.021) = 0.631$

whereas the mean for far transfer items across all curriculum groups is computed as:

$(1)(0.608) + (0)(0.083) + (0)(0.079) + (-1)(0.023) + (0)(0)(0.011) + (0)(0)(0.021) = 0.585$

Computing the mean for a particular curriculum group for either only the far transfer items or only the near transform items simply requires identifying the correct set of coefficients. For example, the mean of the near transfer items for the CAT-UMINN group is given by:

$(1)(0.608) + (1)(0.083) + (0)(0.079) + (1)(0.023) + (1)(1)(0.011) + (0)(0)(0.021) = 0.725$

whereas the mean of the far transfer items for the CAT-REPL group is computed as:

$(1)(0.608) + (-1)(0.083) + (1)(0.079) + (-1)(0.023) + (-1)(-1)(0.011) + (1)(-1)(0.021) = 0.571$

which is within rounding error of the value reported in Table 4 (0.570).